\documentclass{ws-mpla}
\pdfoutput=1
\usepackage{amssymb}
\newcommand{\half}{\frac{1}{2}}

\newcommand{\frth}{\frac{1}{4}}
\newcommand{\sxth}{\frac{1}{6}}
\usepackage[dvipsnames]{color}
\newcommand{\revision}[1]{{           {#1}}} 
\begin{document}
\markboth{R. K. Nesbet}{Conformal Higgs model: gauge fields can produce 
a 125GeV resonance} 
\catchline{}{}{}{}{}
\title{CONFORMAL HIGGS MODEL: GAUGE FIELDS CAN PRODUCE A 125 GEV
RESONANCE}
\author{\footnotesize ROBERT K. NESBET}
\address{IBM Almaden Research Center,
650 Harry Road, San Jose, CA 95120, USA\\
rkn@earthlink.net}
\maketitle
\begin{history}
\received{received date}
\revised{revised date}
\end{history}
\date{\today}
\begin{abstract}
Recent cosmological observations and compatible theory offer an
understanding of long-mysterious dark matter and dark energy.  The
postulate of universal conformal local Weyl scaling symmetry, without 
dark matter, modifies action integrals for both Einstein-Hilbert
gravitation and the Higgs scalar field by gravitational terms.  
Conformal theory accounts both for observed excessive external 
galactic orbital velocities and for accelerating cosmic expansion.
SU(2) symmetry-breaking is retained by the conformal scalar field, 
which does not produce a massive Higgs boson, requiring an alternative
explanation of the observed LHC 125GeV resonance.  
Conformal theory is shown here to be compatible with 
a massive neutral particle or resonance $W_2$ at 125GeV,
described as binary scalars $g_{\mu\nu}W_-^\mu W_+^\nu$ and
$g_{\mu\nu}Z^{\mu*}Z^\nu$ interacting strongly via quark exchange.
Decay modes would be consistent with those observed at LHC.
Massless scalar field $\Phi$ is dressed by the $W_2$ field to 
produce Higgs Lagrangian term $\lambda(\Phi^\dag\Phi)^2$ with the 
empirical value of $\lambda$ known from astrophysics.
\end{abstract}
\ccode{PACS Nos.: 04.20.Cv, 14.80.Gt, 98.80.-k} 

\section{Introduction}
\par In the currently accepted $\Lambda$CDM paradigm for cosmology,
gravitational phenomena not explained by general relativity
as formulated by Einstein are attributed to cold dark matter.
Dark energy $\Lambda$ remains without an explanation. The search for
tangible dark matter has continued for many years\cite{SAN10},
without result.
\par Consideration of an alternative paradigm is motivated by this
situation.  An alternative postulate is that of universal conformal
symmetry\cite{NES13}, requiring local Weyl scaling covariance
\cite{WEY18,MAN06} for all massless elementary physical fields, without 
dark matter. Conformal symmetry, valid for fermion and gauge boson 
fields\cite{DEW64},is extended to both the metric tensor field of general relativity and the Higgs scalar field of elementary-particle 
theory\cite{HIG64,PAS95,CAG98}, with no novel elementary fields. 
Lagrangian densities of both conformal fields include gravitational terms.  Neither model requires dark matter. 
\revision{
Conformal gravity(CG) \cite{MAK89,MAN90,MAN91,MAN06,MAN12,OCM18} 
fits rotation data for 138 galaxies.  The conformal Higgs model (CHM) 
\cite{NESM1,NESM2,NESM3,NESM5} fits observed accelerating 
Hubble expansion for redshifts $z\leq 1$ (7.33 Gyr) accurately with one 
free constant parameter\cite{NESM2}. 
\par The CHM evaluates parameters of the Higgs model using observed 
empirical data from cosmology and electroweak particle physics.  The
implied parameter values are inconsistent with a massive Higgs 
particle\cite{NESM2,NESM5},
requiring an alternative explanation of the observed LHC 125GeV 
resonance\cite{DJV12,ATL12,CMS12}.  The CHM, extended here to include 
charged gauge fields $W^\pm_\mu$, is shown to be consistent with a
novel neutral 125 GeV resonance.    
}
\par The Higgs model of electroweak physics postulates a
scalar field $\Phi$\cite{HIG64,PAS95,CAG98} whose classical field 
equation has an exact stable solution of finite amplitude. 
Gauge symmetry defines covariant derivatives of $\Phi$ that couple it 
to neutral $Z_\mu$ and charged $W^\pm_\mu$ gauge boson fields.
Gauge boson masses result from scalar and gauge boson field 
equations\cite{HIG64,CAG98} evaluated at a semiclassical level. 
Electroweak masses depend only on the existence of a 
spacetime solution of the Higgs scalar field equation
with nonvanishing  spacetime amplitude. The 
induced masses do not require a stable fluctuation of the scalar field,
the usual definition of a massive Higgs boson\cite{CAG98}. 
\par The Higgs model postulates incremental scalar field Lagrangian 
density\cite{CAG98}  
$\Delta{\cal L}_\Phi=w^2\Phi^\dag\Phi-\lambda(\Phi^\dag\Phi)^2$.
The conformal Higgs model\cite{NESM1} acquires an additional 
term\cite{MAN06}
$-\sxth R \Phi^\dag\Phi$ with gravitational Ricci scalar 
$R=g_{\mu\nu}R^{\mu\nu}$. The modified Lagrangian, defined for 
neutral gauge field $Z_\mu$\cite{NESM2},
determine $w^2$, which becomes a cosmological constant
in the conformal Friedmann cosmic evolution equation\cite{NESM1}. 
Lagrangian term $w^2\Phi^\dag\Phi$ is due to induced neutral $Z_\mu$ field\cite{NESM2}, which dresses the bare scalar field.
\revision{  
Finite $w^2$ breaks conformal symmetry, but does not 
determine a Higgs mass.
\par Empirical $\lambda<0$ is shown here to result from 
dressing bare $\Phi$ by an induced neutral scalar field $W_2$,
defined as a strongly interacting pair of gauge bosons. 
Empirical data implies estimated $\lambda$ consistent with a spinless 
particle $W_2$ of mass 125GeV. This might account for the observed  
125GeV resonance\cite{DJV12,ATL12,CMS12}. The model of neutral $W_2$ 
proposed here does not conflict with other implications of electroweak 
theory.  The Higgs mechanism is retained by the massless Higgs scalar
field of finite amplitude.  Parameters $w^2$ and $\lambda$, required
for nonzero $\Phi^\dag\Phi$, both result from the $U(1)\times SU(2)$ covariant derivative of the Higgs field, which produces finite 
classical source density for two neutral fields.  Neutral $Z^\mu$ 
dresses field $\Phi$ to produce Lagrangian term $w^2\Phi^\dag\Phi$.  Neutral $W_2$ dresses field $\Phi$ to produce 
$-\lambda(\Phi^\dag\Phi)^2$.
}
\par Conformal gravity\cite{MAN06,MAN12} has recently been applied to 
fit anomalous galactic rotation data for 138 
galaxies\cite{MAO11,MAO12,OAM12},
without dark matter. Formal objections\cite{FLA06,BAV09,YOO13} have been 
discussed and resolved in detail\cite{MAN07,NESM5,NESM7}. 
McGaugh et al\cite{MLS16} show for 153 
galaxies that observed radial acceleration $a$ is effectively a 
universal function of the classical Newtonian acceleration $a_N$, 
computed for the observed baryonic distribution. 
Such a universal correlation function, $a(a_N)=\nu(a_N/a_0)a_N$, 
is a basic postulate of MOND\cite{MIL83,FAM12}.
This removes uncertainty in earlier studies due to 
adjustment of mass-to-light ratios for individual galaxies.
\revision{
This conclusion is disputed\cite{OCM18} by a purely CG fit to the  
same data. Resolution requires more accurate data at large radii. 
} 
Together with the CHM and the depleted halo model of
galactic halos\cite{NESM3}, without dark matter, CG implies a 
similar correlation function if Mannheim nonclassical acceleration 
$\gamma$ is mass-independent\cite{NESM7}.  This strongly suggests,
for an isolated galaxy, that dark matter can be eliminated and dark
energy explained by consistent conformal 
theory\cite{MAN06,MAN12,NES13,NESM7,NESM5}.

\section{Review of theory}
\par Variational theory for fields in general relativity is a
straightforward generalization of classical field theory\cite{NES03}.
Given scalar Lagrangian density ${\cal L}$, action integral
$I=\int d^4x \sqrt{-g} {\cal L}$ is required to be stationary for
all differentiable field variations, subject to appropriate boundary
conditions.  The determinant of metric tensor $g_{\mu\nu}$ is denoted
here by $g$. Riemannian metric covariant derivatives ${\cal D}_\lambda$
are defined such that\cite{MAN06} ${\cal D}_\lambda g_{\mu\nu}=0$.
\par Conformal symmetry is defined by invariance of action integral
$I=\int d^4x\sqrt{-g}{\cal L}$ under local Weyl scaling\cite{WEY18}, 
such that $g_{\mu\nu}(x)\to g_{\mu\nu}(x)\Omega^2(x)$ for arbitrary
real differentiable $\Omega(x)$, with fixed coordinates $x^\mu$.
For any Riemannian tensor, $T(x)\to \Omega^d(x)T(x)+{\cal R}(x)$
defines weight $d[T]$ and residue ${\cal R}[T]$.  $d[\Phi]=-1$ for a
scalar field.  Conformal Lagrangian density ${\cal L}$ must have weight
$d[{\cal L}]=-4$ and residue ${\cal R}[{\cal L}]=0$,
up to a 4-divergence\cite{MAN06}.
\par Gravitational field equations are determined by metric
functional derivative
$X^{\mu\nu}= \frac{1}{\sqrt{-g}}\frac{\delta I}{\delta g_{\mu\nu}}$.
Any scalar ${\cal L}_a$ determines energy-momentum tensor
$\Theta_a^{\mu\nu}=-2X_a^{\mu\nu}$,
evaluated for a solution of the coupled field equations.
Generalized Einstein equation $\sum_aX_a^{\mu\nu}=0$ is expressed as
$X_g^{\mu\nu}=\half\sum_{a\neq g}\Theta_a^{\mu\nu}$.  Hence summed trace
$\sum_ag_{\mu\nu}X_a^{\mu\nu}$ vanishes for exact field solutions. 
\revision{Concurrent solution of exact equations for both Higgs field 
$\phi_0(t)$ and Friedmann scale $a(t)$ is needed for accurate extension  
to large redshifts $z$.  The derivation\cite{NESM5} of $a(t)$
for $z\leq 1$ and an earlier derivation\cite{NESM1} for 
$z\leq 1090$(CMB) treat $\dot{\phi}/\phi$ as a constant parameter.}

\section{Conformal scalar field}
\par The fundamental postulate that all primitive fields have conformal
Weyl scaling symmetry is satisfied by spinor and gauge
fields\cite{DEW64}, but not by the scalar field of the conventional
Higgs model\cite{CAG98}.
A conformally invariant action integral is defined for complex SU(2)
doublet scalar field $\Phi$ by Lagrangian density\cite{MAN06,NESM1}
\begin{eqnarray}\label{LPhi}
{\cal L}_\Phi=
 (\partial_\mu\Phi)^\dag\partial^\mu\Phi-\sxth R\Phi^\dag\Phi
 -\lambda(\Phi^\dag\Phi)^2,
\end{eqnarray}
where $R$ is the gravitational Ricci scalar. The Higgs
model\cite{CAG98} postulates incremental Lagrangian density
$\Delta{\cal L}_\Phi$, which adds term $w^2\Phi^\dag\Phi$ to
${\cal L}_\Phi$.  Because this $w^2$ term breaks conformal symmetry,
universal conformal symmetry requires it to be produced dynamically.
\par The conformal scalar field equation including 
parametrized $\Delta{\cal L}_\Phi$ is\cite{MAN06,NESM1} 
\begin{eqnarray}\label{Phieq}
\frac{1}{\sqrt{-g}}\partial_\mu(\sqrt{-g}\partial^\mu\Phi)=
 (-\sxth R+w^2-2\lambda\Phi^\dag\Phi)\Phi.
\end{eqnarray}
Neglecting second-order time derivative terms for $\Phi$,
$\Phi^\dag\Phi=\phi_0^2=(w^2-\sxth R)/2\lambda$
generalizes the Higgs construction if this ratio is positive.
$w^2<\sxth R$, which implies\cite{NESM1} $\lambda<0$.
\par Because Ricci scalar $R$ varies on a cosmological time scale,
it induces an extremely weak time dependence of $\phi_0^2$, which in
turn produces source current densities for the gauge fields.
The resulting coupled semiclassical field equations\cite{NESM2}
determine nonvanishing but extremely small parameter $w^2$, in agreement
with the cosmological constant deduced from Hubble expansion data. This
argument depends only on squared magnitudes of quantum field amplitudes.
\revision{
Conformal theory is shown here to determine
biquadratic term $\lambda(\Phi^\dag\Phi)^2$
by a similar semiclassical argument.
}

\section{Gauge fields and Higgs parameters}
\par In standard Higgs theory, it is assumed that charged component 
$\Phi_+$ of the postulated SU(2) doublet scalar field vanishes 
identically, while $\Phi^\dag\Phi=\phi_0^2$ is a spacetime constant.
In conformal theory, Ricci scalar $R$ causes $\phi_0$ to vary
on a cosmological time scale\cite{NESM2}.
Extending a derivation for neutral $Z^\mu$\cite{NESM1,NESM2}
verifies standard mass formulae
$m_Z^2=\half g_z^2\phi_0^2, m_W^2=\half g_w^2\phi_0^2, m_A^2=0$.
$\Delta{\cal L}$ implies source current density 
$J_Z^0=-ig_z\frac{{\dot\phi}_0}{\phi_0}\Phi^\dag\Phi$ and
$Z^0=J_Z^0/m_Z^2$, which determines Higgs parameter 
$w^2=\frth g_z^2 Z_\mu^*Z^\mu=(\frac{{\dot\phi}}{\phi})^2$.  
\par From the covariant derivative\cite{CAG98} of $\Phi$, with
$\Phi_+\equiv 0$, 
\begin{eqnarray}
\Delta{\cal L}= 
(-\frac{ig_w}{\sqrt{2}}W_{-\mu}\Phi^\dag) 
 (\frac{ig_w}{\sqrt{2}}W_+^\mu\Phi) 
\nonumber \\
+((\partial_\mu+\frac{ig_z}{2}Z_\mu^*)\Phi^\dag)
 (\partial^\mu-\frac{ig_z}{2}Z^\mu)\Phi
-\partial_\mu\Phi^\dag \partial^\mu\Phi
\nonumber \\
= (\frth g_w^2(W^*_{+\mu}W_+^\mu+W^*_{-\mu}W_-^\mu)
+\frth g_z^2 Z_\mu^*Z^\mu)\Phi^\dag\Phi
\nonumber \\
-\partial_\mu\Phi^\dag(\half ig_z Z^\mu\Phi)
+(\half ig_z Z_\mu^*\Phi^\dag)\partial^\mu\Phi.
\end{eqnarray}
\par $W_{\mu-}W_+^\mu=
\half(W^*_{+\mu}W_+^\mu+W^*_{-\mu}W_-^\mu)$ 
is shown here to contribute to $-\lambda(\Phi^\dag\Phi)^2$. 
Although an independent field $W_\pm^\mu$ would violate charge
neutrality, there is no contradiction in treating neutral composite
scalar $WW=g_{\mu\nu}W_-^\mu W_+^\nu$ as an independent field or 
particle, in analogy to atoms, molecules, and nuclei.
Bare $WW$ must interact strongly with corresponding neutral scalar
field $ZZ=g_{\mu\nu}Z^{\mu*}Z^\nu$, through exchange of quarks and
leptons.  Assuming that the interacting bare fields produce relatively 
stable $W_2=WW\cos\theta_x+ZZ\sin\theta_x$ and complementary
resonance $Z_2=-WW\sin\theta_x+ZZ\cos\theta_x$, $W_2$ can dress the 
bare $\Phi$ field while maintaining charge neutrality.  This will be 
shown here to determine Higgs parameter $\lambda$.
\par $\Delta{\cal L}$ does not determine an explicit source 
density for field $WW$.  However, simultaneous creation of
fields $Z_\mu^*,Z^\mu$ can produce the dressed field $W_2$.
The rate of simultaneous creation is the product 
of independent rates, expressed by
$J_{ZZ}/\Phi^\dag\Phi=
(J_{Z\mu}^*/\Phi^\dag\Phi)(J_Z^\mu/\Phi^\dag\Phi)$.
This implies $J_{ZZ}\Phi^\dag\Phi=J_{Z\mu}^* J_Z^\mu=
g_z^2 (\frac{{\dot\phi}_0}{\phi_0})^2(\Phi^\dag\Phi)^2$.
Assuming a $W_2$ field containing 
linear combination $WW\cos\theta_x+ZZ\sin\theta_x$, the effective
source current is $J_{W_2}=J_{ZZ}\sin\theta_x$.
Neglecting derivatives in an effective Klein-Gordon equation,
the induced field amplitude is $W_2=J_{W_2}/m^2_{W_2}$. 
Given $WW=W_2\cos\theta_x-Z_2\sin\theta_x$,  
$W_2$ projects onto $WW$ with factor $\cos\theta_x$.
Term $\half g_w^2 WW\Phi^\dag\Phi$ in $\Delta{\cal L}$ becomes
$\frth g_w^2\sin2\theta_x J_{ZZ}\Phi^\dag\Phi/m^2_{W_2}=
-\lambda(\Phi^\dag\Phi)^2$,
where dimensionless $\lambda=-\frth g_w^2g_z^2
\sin2\theta_x(\frac{{\dot\phi_0}}{\phi_0})^2\hbar^2/m^2_{W_2}c^4$.
\revision{
\par If $\lambda$ is constant\cite{NESM5}, parameter 
$\frac{{\dot\phi_0}}{\phi_0}=-2.731H_0$,
where Hubble constant\cite{PLC15}
$H_0=67.66km/s/Mpc=2.197\times 10^{-18}/s$.
The $W_2$ mass must be consistent with empirical\cite{NESM5}
$\lambda\simeq -10^{-88}$.
}

\section{$W_2$ particle and $Z_2$ resonance}
\revision{
\par Conformal theory evaluates otherwise undetermined Higgs field
parameters from accurately fitted observed cosmic expansion and galactic
rotation. The resulting empirical parameter values remain consistent
with the standard electroweak model, requiring only spontaneous 
generation of finite neutral Higgs field amplitude $\phi_0$, but
are incompatible with existence of a massive Higgs scalar particle.
Hence the observed LHC 125GeV resonance\cite{DJV12,ATL12,CMS12} 
requires an alternative explanation. 
}
\par A model Hamiltonian matrix can be defined in which indices 0,1 
refer respectively to bare neutral scalar states 
$WW=g_{\mu\nu}W_-^\mu W_+^\nu$, $ZZ=g_{\mu\nu}Z^{\mu*} Z^\nu$. Assumed 
diagonal elements are $H_{00}=2m_W=160GeV$, $H_{11}=2m_Z=182GeV$,
for empirical masses $m_W$ and $m_Z$.
Intermediate quark and lepton states define a large complementary matrix
${\tilde H}$ indexed by $i,j\neq 0,1$, with eigenvalues $\epsilon_i$, 
and off-diagonal elements ${\tilde A}_{i0},{\tilde A}_{i1}$.  
${\tilde H}$ determines energy-dependent increments in a
$2\times 2$ reduced matrix 
\begin{eqnarray}
 H_{ab}-\mu_{ab}=H_{ab}-\sum_{i \neq 0,1}
{\tilde A}^\dag_{ai}(\epsilon_i-\epsilon)^{-1}{\tilde A}_{ib} .
\end{eqnarray}
\par $H_{01}-\mu_{01}=(WW|H_{red}|ZZ)$ corresponds to Feynman diagrams
for quark and lepton exchange.  The most massive and presumably most
strongly coupled intermediate field that interacts directly would be
tetraquark $T=t{\bar b}b{\bar t}$,
whose mass is estimated as $\epsilon_T=350$GeV.  
A simplified estimate of $W_2$ energy is obtained by restricting 
intermediate states to the three color-indexed tetraquark states
$T=t{\bar b}b{\bar t}$, and assuming elements 
${\tilde A}_{T0},{\tilde A}_{T1}$ of equal magnitude $\alpha/\sqrt{3}$.
For the reduced $2\times2$ matrix, matrix increments
$\mu_{ab}\simeq\mu(\epsilon)=
\frac{\alpha^2}{\epsilon_T-\epsilon}$
are all defined by a single parameter $\alpha^2$.
Secular equation 
\begin{eqnarray}
 (2m_W-\mu(\epsilon)-\epsilon)(2m_Z-\mu(\epsilon)-\epsilon)=
\mu^2(\epsilon) 
\end{eqnarray}
is to be solved for two eigenvalues $\epsilon=E_0,E_1$.
\par It is found that identifying the model diboson $W_2$ with the
recently observed LHC 125GeV resonance\cite{DJV12,ATL12,CMS12}
predicts the empirical value of Higgs parameter $\lambda$.
Setting $E_0=125GeV=0.8644\times 10^{44}\hbar H_0$ 
for the $W_2$ state, dominated by the bare 
$WW$ field, determines parameters $\alpha^2=4878$ GeV$^2$,
$\mu(E_0)=21.68$GeV and $\tan\theta_x=0.6138$.
Using $\alpha^2$ determined by $E_0$, the present model predicts 
$E_1=173$GeV, with $\mu(E_1)=27.62$GeV.  This higher eigenvalue 
is the energy of a resonance $Z_2$ dominated by the bare $ZZ$ field.
$Z_2$ decay into bare $WW$, two free charged gauge bosons, is allowed
by energy conservation, but not into bare $ZZ$.  Composite field $W_2$ 
cannot decay spontaneously into either $WW$ or $ZZ$.
\par Identifying $E_0$ with the observed 125GeV resonance, and using 
$g_w=0.6312$ and $g_z=0.7165$ with computed $\tan\theta_x=0.6138$,
the implied value of Higgs parameter 
$\lambda=-\frth g_w^2g_z^2
\sin2\theta_x(\frac{{\dot\phi}_0}{\phi_0})^2\hbar^2/m^2_{W_2}c^4
=-0.455\times10^{-88}$ is consistent with its empirical 
value\cite{NESM5} $\lambda\simeq -10^{-88}$.

\section{Conclusions and implications}
\revision{
\par Postulated universal conformal symmetry modifies both
general relativity and the Higgs scalar field model.
Higgs parameter $w^2$ and all mass terms break
conformal symmetry and must be generated dynamically.  The conformal 
Higgs model remains valid for gauge boson masses, but the negative
sign of parameter $\lambda$ implied by cosmological data\cite{NESM5} 
precludes a massive particle as a Higgs scalar field fluctuation.
}  
\par The coupled semiclassical field equations of conformal theory 
imply a very small but nonvanishing source density for the neutral 
gauge field $Z_\mu$\cite{NES13,NESM2}.
This results from the cosmological time dependence of gravitational
Ricci scalar $R$ in the conformal scalar field Lagrangian
density.  Bare Higgs scalar $\Phi$ is dressed by a nonvanishing
neutral gauge field, producing parameter $w^2$ of the correct empirical 
magnitude for dark energy density\cite{NESM1,NESM2}. 
\par Agreement with empirical data is extended here
to Higgs parameter $\lambda$.  Preserving charge neutrality,
double excitation of the $Z^\mu$ field induces a previously unknown
field $W_2$, based on strongly interacting bare fields
$WW=W_{\mu-}W_+^\mu$ and $ZZ=Z^*_\mu Z^\mu$, which dresses bare
scalar field $\Phi$.  Implied parameter $\lambda$ has empirically 
correct sign and magnitude if scalar field $W_2$ is identified 
with the recently observed 125GeV resonance\cite{DJV12,ATL12,CMS12}.
\par Explanation of Higgs parameter $\lambda$ makes it possible to 
carry the modified Friedmann cosmic evolution equation\cite{NESM1}
back to the big-bang epoch, with time-dependent parameters.
The reversed sign of the conformal gravitational constant in uniform,
isotropic geometry\cite{MAN06,NES13} implies rapid expansion due to
primordial mass and radiation density.  Time variation 
of conformal Higgs amplitude $\phi_0$ is relevant to nucleosynthesis, 
because it directly affects the Fermi $\beta$-decay constant.
\par Dynamical models of galactic clusters should be revised to take
into account the non-Newtonian gravitational effects of conformal
theory\cite{NES13,NESM3}.  It cannot yet be concluded that dark matter 
is needed to explain galactic evolution.
\par The conformal Higgs model, backed by well-determined cosmological
data, does not imply a massive Higgs particle, but supports the
alternative interpretation proposed here of the observed 125GeV
resonance.  If Higgs parameter 
$w^2$ were large enough to produce the observed 125GeV resonance, the 
conformal model would imply dark energy density large enough to blow 
the universe apart long before the present epoch.  This implication is 
removed by the present theory.  Moreover, without a stable fluctuation,
the symmetry-breaking Higgs $\Phi$ field does not produce large
zero-point energy that must somehow be suppressed. 
An extremely small scale parameter
$\frac{{\dot\phi}_0}{\phi_0}$, unique to conformal theory, 
relates cosmology to electroweak physics.
\par The Higgs scalar field breaks gauge symmetry and produces gauge
boson mass through its finite spacetime amplitude.
The standard electroweak model\cite{PAS95,CAG98} attributes fermion 
mass to direct coupling with the Higgs scalar field.
This mechanism is preserved by the conformal Higgs field amplitude.


\end{document}